\documentclass{andromedaone}       

\usepackage{bbm}
\usepackage{amsmath, amssymb}
\usepackage{xspace}
\usepackage{url}
\usepackage{hyperref}
\usepackage{nicefrac}

\newcommand{\Z}[1]{\ensuremath{\mathbbm{Z}_{#1}}}
\newcommand{\SO}[1]{\ensuremath{\mathrm{SO}(#1)}}
\newcommand{\SU}[1]{\ensuremath{\mathrm{SU}(#1)}}
\newcommand{\U}[1]{\ensuremath{\mathrm{U}(#1)}}
\newcommand{\I}{\mathrm{i}}
\newcommand{\rep}[1]{\ensuremath\boldsymbol{#1}}

\newcommand{\Id}{\mathbbm{1}}
\newcommand{\CP}{\ensuremath{\mathcal{CP}}\xspace}

\journal{BSM}
\vol{2021}
\jyear{Egypt}
\pages{Zewail City} 

\received{xx May 2021}
\published{xx xxx 2021}

\def\be{\begin{equation}}
\def\ee{\end{equation}}
\def\bea{\begin{eqnarray}}
\def\eea{\end{eqnarray}}

\begin{document}

\title{Flavor and \CP from String Theory}

\author{Hans Peter Nilles,\auno{1} Sa\'ul Ramos-S\'anchez\auno{2} and Patrick K.S. Vaudrevange\auno{3}}
\address{$^1$Bethe Zentrum f\"ur Theoretische Physik, Universit\"at Bonn, 
Nussallee 12, 53115 Bonn, Germany}
\address{$^2$Instituto de F\'isica, Universidad Nacional Aut\'onoma de M\'exico, POB 20-364, Cd.Mx. 01000, M\'exico}
\address{$^3$Physik Department T75, Technische Universit\"at M\"unchen, James-Franck-Stra\ss e 1, 85748 Garching, Germany}

\begin{abstract}
Modular transformations of string theory are shown to play a crucial role in the discussion of 
discrete flavor symmetries in the Standard Model. They include \CP transformations and provide a 
unification of \CP with traditional flavor symmetries within the framework of the ``eclectic 
flavor'' scheme. The unified flavor group is non-universal in moduli space and exhibits the 
phenomenon of ``Local Flavor Unification'', where different sectors of the theory (like quarks and 
leptons) can be subject to different flavor structures.\\
\end{abstract}

\maketitle

\begin{keyword}
flavor symmetry\sep string theory constructions\sep modular transformations\sep
eclectic flavor group\sep local flavor unification
\end{keyword}

\section{Outline}
\begin{flushright}
\vspace{-8.6cm}\normalsize{TUM-HEP 1336/21}\vspace{8cm}
\end{flushright}

We shall discuss the world of flavor from a top-down point of view. The topics include
\begin{itemize}
\item {\bf Traditional flavor symmetries}
\item {\bf Modular flavor symmetries}
\item {\bf \boldmath Natural appearance of a \CP symmetry\unboldmath}
\item {\bf The concept of eclectic flavor symmetry}
\item {\bf A unified picture of quark- and lepton-flavor: flavor groups localized in moduli space}
\item {\bf \boldmath Spontaneous breakdown of flavor and \CP symmetry\unboldmath}
\item {\bf Comparison of top-down and bottom-up constructions}
\item {\bf Conclusion and outlook}
\end{itemize}
In this talk  (at BSM2021, Zewail City, Egypt by HPN)
we are reporting about joint work in various combinations 
with Alexander Baur, Moritz Kade and Andreas Trautner.

\section{Discrete Flavor Symmetries}
\subsection{\bf Bottom-up}

Most of the parameters of the $\SU{3}\times\SU{2}\times\U{1}$ Standard Model of particle physics 
concerns the flavor sector: masses and mixing angles of quarks and leptons. Their origin is not yet 
understood. There are many fits to the data from a bottom-up perspective (see 
e.g.~\cite{Feruglio:2019ktm}) that postulate discrete flavor symmetries (like e.g. 
$S_3, A_4, S_4, A_5, \Delta(27), \Delta(54)$, etc.~\cite{Ishimori:2010au}) and choose specific 
representations of these groups for the Standard Model fermions. The data seem to require different 
models for quark- and lepton-sector with qualitative different flavor structures. While in the 
quark sector we observe small mixing angles, this is different in the lepton sector. Moreover, 
flavor symmetries hold only approximately and have to be broken spontaneously. This requires the 
introduction of so-called flavon fields whose vacuum expectation values are responsible for this 
breakdown. With the choice of these flavon fields one introduces many additional parameters. With 
specific choices of these parameters (and the discrete flavor group as well as the relevant 
representations) the model building from the bottom-up perspective leads to many different 
successful fits to masses and mixing angles of matter fields. Still there is not yet a clear 
preference for a given class of models and we seem to need a top-down explanation of the flavor 
puzzle.
                      
\subsection{\bf Top-down}

Such a top-down explanation might come from string theory. Discrete symmetries have their origin in 
the symmetries of compact extra dimensions as well as specific string theory selection rules (that 
arise from conformal symmetry on the string worldsheet)~\cite{Kobayashi:2006wq}. As an 
illustration, we consider two-dimensional orbifold compactifications of heterotic string 
theory~\cite{Dixon:1985jw,Dixon:1986jc,Ibanez:1986tp}. 
They are directly relevant for flavor symmetries of 
six-dimensional string compactifications with an elliptic fibration. They provide the chiral 
spectrum of quarks and leptons at low energies within the $\SU{3}\times\SU{2}\times\U{1}$ Standard 
Model~\cite{Buchmuller:2005jr,Lebedev:2006kn,Nilles:2008gq}. In addition, they display abundant 
discrete symmetries for flavor physics that might allow a connection to the existing bottom-up 
constructions~\cite{Kobayashi:2006wq,Olguin-Trejo:2018wpw,Baur:2019kwi}.

As we shall see, the string theory picture predicts the existence of
\begin{itemize}
\item {\bf traditional flavor symmetries that are universal in moduli space,}
\item {\bf modular flavor symmetries which are non-universal in moduli space,}
\item{\bf  a natural candidate for a \CP symmetry.}
\end{itemize}

The non-universality of modular flavor symmetries in moduli space leads to the phenomenon of ``Local 
Flavor Unification'' at specific points (or higher-dimensional sub-regions) in moduli 
space~\cite{Baur:2019kwi}. This might allow the explanation of the different flavor structures in 
the quark- and lepton-sector of the Standard Model. The spontaneous breakdown of flavor and \CP 
symmetries can be understood as a motion in moduli space (away from these specific points or 
subregions). 

\section{Flavor symmetry from string theory}

\begin{figure}[b!]
\centering
\includegraphics[height=2.3in]{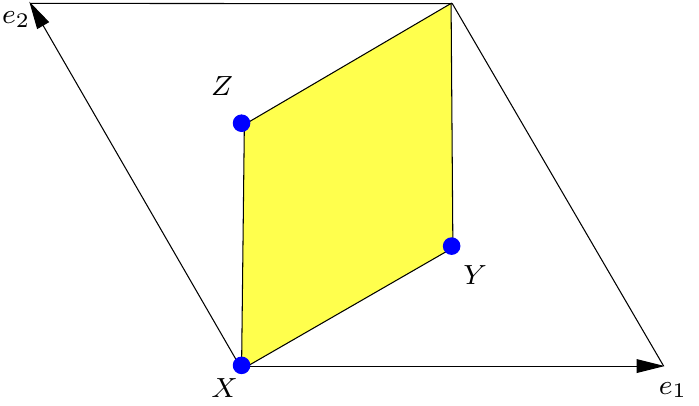}
\caption{\Z3 orbifold (yellow-shaded region) with the three fixed points $X,Y,Z$. Twisted strings
are located at these fixed points.}
\label{fig:T2/Z3}
\end{figure}

Discrete symmetries can arise from geometry and string selection rules. As an illustrative example, 
let us consider the two-dimensional \Z3 orbifold $\mathbbm{T}^2/\Z{3}$ (see Figure~\ref{fig:T2/Z3}). 
The lattice vectors $e_1$ and $e_2$ have the same length and are separated by an angle of 120 
degrees (to allow for the \Z3 twist). The spectrum of the string theory contains untwisted modes, 
winding modes as well as twisted modes. The latter are located at the fixed points $X, Y, Z$ of the 
orbifold twist. In this picture, we obtain an $S_3$ symmetry from the interchange of the fixed 
points. Orbifold selection rules (from the underlying conformal field theory on the string 
worldsheet) add a $\Z{3}\times\Z{3}$ symmetry. The multiplicative closure of $S_3$ and 
$\Z{3}\times\Z{3}$ leads to a group with 54 elements called $\Delta(54)$, a discrete non-Abelian 
subgroup of \SU3, and could e.g.\ describe the flavor symmetry of three families of 
quarks~\cite{Carballo-Perez:2016ooy}. The twisted fields transform as a 3-dimensional
representation of $\Delta(54)$. 

We see that even such simple systems lead to sizeable flavor groups. A full determination of the 
flavor symmetries in six compact dimensions could then lead to a complicated flavor structure that 
requires a general method to classify these flavor symmetries. Such a mechanism has been identified 
via the outer automorphisms~\cite{Baur:2019kwi,Baur:2019iai} of the so-called Narain space 
group~\cite{Narain:1985jj,Narain:1986am,Narain:1986qm,GrootNibbelink:2017usl}. Here, we shall not 
be able to give a full derivation of this fact but illustrate it in our examples (for a 
$D=2$-dimensional torus and for an orbifold).

On a string there are $D$ right-moving and $D$ left-moving degrees of freedom: 
$Y=(y_\mathrm{R}, y_\mathrm{L})$. A geometrical compactification of strings on a $D$-dimensional 
torus amounts to a compactification of $Y$ on a $2D$-dimensional auxiliary torus, obtained by 
demanding the boundary condition
$$
Y ~=~
\begin{pmatrix}
      y_\mathrm{R} \\      
      y_\mathrm{L} 
\end{pmatrix}
~\sim~ Y + E\, \hat N ~=~
\begin{pmatrix}
      y_\mathrm{R} \\      
      y_\mathrm{L} 
\end{pmatrix}
+ E
\begin{pmatrix}
      n \\      
      m
\end{pmatrix},
\quad n,m\in\Z{}^D\,,
$$
that defines a so-called $2D$ Narain lattice (four-dimensional for the $D=2$-torus). It includes 
the string's winding and Kaluza--Klein quantum numbers $n$ and $m$, respectively. The Narain 
vielbein matrix $E$ depends on the moduli of the torus: radii, angles and antisymmetric tensor 
fields. It is important to note that here we do include not only the (momentum) Kaluza--Klein modes 
of the compactified space but also the string winding modes on the nontrivial cycles of the torus. 
In order to arrive at a $D$-dimensional orbifold in the Narain formulation, we now introduce a 
\Z{K} orbifold twist $\Theta$ that leads to the identification
$$
Y ~\sim~ 
\Theta^k\, Y + E\, \hat N, \quad {\rm where}\quad \Theta ~=~
\begin{pmatrix}
  \theta_\mathrm{R} & 0 \\     
  0                 & \theta_\mathrm{L} 
\end{pmatrix}\quad {\rm and} \quad \Theta^K ~=~ \Id_{2D}\,,$$
with $\theta_\mathrm{L}, \theta_\mathrm{R}$ elements of $\SO{D}$ and $k=0,\ldots,K-1$. For a 
symmetric orbifold, we set $\theta_\mathrm{L}=\theta_\mathrm{R}$. The Narain space group, built by 
elements $g=(\Theta^k, E\hat N)$, is then generated by
$$
{\rm the\ twist}\ \ (\Theta, 0)\ \ {\rm and \ shifts}\ \  (1, E_i) \ \ {\rm for} \ \ i = 1, \ldots, 2D\;,
$$
where we do not include roto-translations (that would correspond to off-diagonal terms in $\Theta$).
Flavor symmetries correspond to the outer automorphisms of this
Narain space group. Outer automorphisms map the group to itself but are not elements of the group. 
Note that they also include modular transformations that interchange winding and momentum modes.

\section{Modular transformations in string theory}

\begin{figure}[t!]
\centering
\includegraphics[height=3.5in]{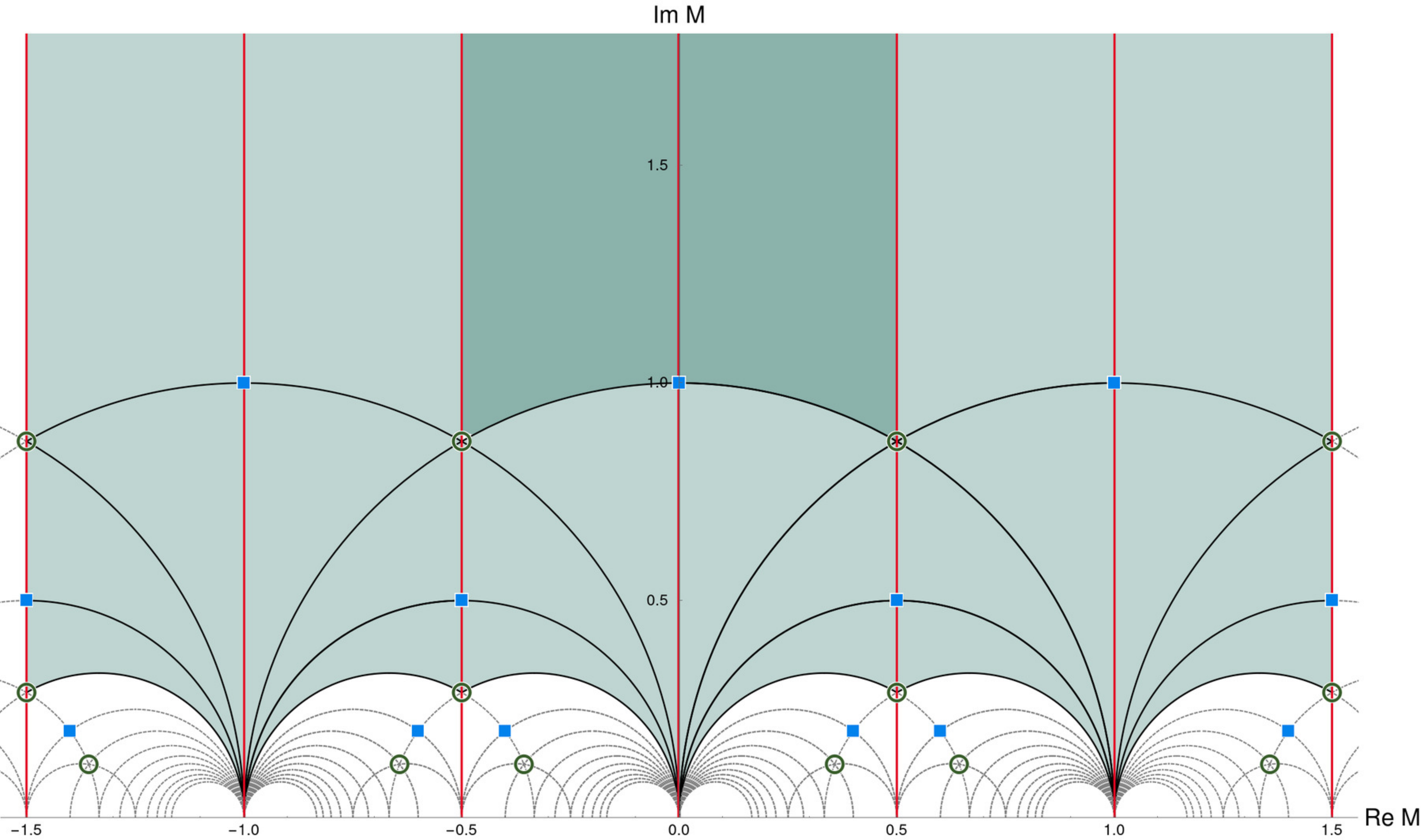}
\caption{The dark shaded region is the fundamental domain of $\mathrm{SL}(2,\Z{})$ (extending to 
$M=\I\infty$). The light shaded region is the fundamental domain of the finite modular group 
$T^\prime$.}
\label{fig:ModuliFundamentalDomain}
\end{figure}

Duality symmetries are frequently found in string theory. One prominent example among them is 
T-duality that exchanges winding and momentum modes. Consider strings on a circle of radius $R$. 
The discrete spectrum of the momentum (Kaluza--Klein) modes is governed by $1/R$, while for the 
winding modes we find a spacing $R$. Winding states become heavy when the radius of the circle 
increases. The T-duality transformation of string theory exchanges 
$$
{\rm winding} ~\mapsto~ {\rm momentum} 
$$
modes and, simultaneously 
$$
R ~\mapsto~ \alpha'/R\;. 
$$
This maps a theory to its T-dual with a self-dual point at 
$$
R^2 ~=~ \alpha^\prime ~=~ 1/M_{\rm string}^2\;,
$$
where $M_{\rm string}^2$ is the string tension. If the string scale is large, then the low energy 
effective field theory describes the momentum modes and the winding states will be heavy. This 
raises the question whether T-duality can be relevant to flavor physics. Before we answer this 
question to the positive, let us generalize the circle compactification to higher-dimensional tori. 
This leads us to modular transformations that still exchange winding and momentum modes and act 
nontrivially on the moduli of the torus. In $D=2$ these transformations are connected to the group 
$\mathrm{SL}(2,\Z{})$ acting on the moduli (K\"ahler and complex structure modulus) of the 
$D=2$-torus. The group $\mathrm{SL}(2,\Z{})$ is generated by two elements
$$
\mathrm{S}, \mathrm{T} \ \ \ \ {\rm with}\ \ \ \   \mathrm{S}^4 ~=~ \Id\,,\qquad \mathrm{S}^2\mathrm{T} ~=~ \mathrm{T}\mathrm{S}^2
\ \ \ \ {\rm and}\ \ \ \  (\mathrm{S}\mathrm{T})^3 ~=~ \Id\;.
$$
For each modular group $\mathrm{SL}(2,\Z{})$, there exists an associated modulus $M$ that 
transforms as
$$\mathrm{S}:\ \ \ M ~\mapsto~ -{1\over M} \ \ \      \ \ \ {\rm and}\ \ \ \ \   \mathrm{T}: \ \ \  M ~\mapsto~ M+1\;.
$$
Further transformations include mirror symmetry (exchanging K\"ahler and complex
structure modulus) as well as the transformation
$$
M ~\mapsto~ -\overline{M}\,,
$$
where $\overline{M}$ denotes the complex conjugate of $M$. The latter turns out to be a universal 
and natural candidate for a \CP symmetry. String dualities give important constraints on the action 
of the theory via the modular group $\mathrm{SL}(2,\Z{})$. Combining $\mathrm{S}$ and $\mathrm{T}$, 
the general transformation $\gamma\in\mathrm{SL}(2,\Z{})$ of the modulus is given by
$$
M ~\stackrel{\gamma}{\mapsto}~ {{a\,M+b}\over{c\,M+d}}
$$
with $ad-bc=1$ and integers $a,b,c,d$. The value of $M$ (originally in the upper complex half 
plane) is then restricted to the fundamental domain, as shown in 
Figure~\ref{fig:ModuliFundamentalDomain}. Matter fields $\phi$ transform similar to modular forms 
of weight $k$
$$
\phi ~\stackrel{\gamma}{\mapsto}~ (c\,M+d)^k \rho(\gamma)\,\phi \quad{\rm for}\quad \gamma ~\in~ \mathrm{SL}(2,\Z{})\;,
$$
where $(c\,M+d)^k$ is known as automorphy factor and $\rho(\gamma)$ is a representation of 
$\gamma$. It is important to realize that $(\rho(\mathrm{T}))^N = \Id$ even though 
$\mathrm{T}^N\neq\Id$ such that $\rho(\gamma)$ generates a finite group, the so-called finite 
modular group, as we will discuss shortly in more detail. Among others, the modular weights $k$ of 
the fields are important ingredients for flavor model building. The Yukawa couplings are modular 
forms and transform properly (with the modular weights determined by the theory under consideration). 
Then, in a supersymmetric theory, the combination $G=K + \log({\mathcal{W}\overline{\mathcal{W}}})$ 
of K\"ahler potential $K$ and superpotential $\mathcal{W}$ must be invariant under modular 
transformations.

Let us now consider the potential relevance of modular symmetries for flavor physics. Again, we 
would like to illustrate this with the $\mathbbm{T}^2/\Z{3}$ orbifold (see Figure~\ref{fig:T2/Z3}). 
On this orbifold some of the moduli are frozen. Here, the length of the lattice vectors $e_1$ 
and $e_2$ are equal with an angle of 120 degrees between them. This gives restrictions on how 
modular transformations act on the matter fields. For twisted string modes, the coefficients 
$a,b,c,d$ in the transformation
$$
M ~\mapsto~ {{a\,M+b}\over{c\,M+d}}
$$
are defined only modulo 3, which amounts to $N=3$ in $(\rho(\mathrm{T}))^N = \Id$. This constraint 
indicates that for $\mathbbm T^2/\Z3$ a modular transformation $\gamma$ acts trivially on matter 
fields if $\gamma$ belongs to the so-called principal congruence subgroup\footnote{Recall that 
$\Gamma(N)$ is defined abstractly as the infinite subgroup of $\mathrm{SL}(2,\Z{})$, where 
$a,d=1\mod N$ and $b,c=0\mod N$.} $\Gamma(3)\subset\mathrm{SL}(2,\Z{})$. For more general \Z{K} 
orbifolds, other infinite subgroups of $\mathrm{SL}(2,\Z{})$ arise. E.g.\ for $K=2$, the modular 
symmetry of the $\Z{2}$ orbifold is related to $\Gamma(2)$. The quotients 
$\Gamma_N = \mathrm{PSL}(2,\Z{})/\Gamma(N)$ and their double covers 
$\Gamma'_N = \mathrm{SL}(2,\Z{})/\Gamma(N)$ are called finite modular groups, and these are the 
groups that will become relevant as discrete flavor symmetries. In our example, we have $N=3$ and 
the finite modular group is $\Gamma'_3 \cong T'\cong\mathrm{SL}(2,3)$ (the double cover of 
$\Gamma_3 \cong A_4$, the group of even permutations of four objects). The fundamental domain of 
$\Gamma'_3 \cong T'$ is bigger than that of $\mathrm{SL}(2,\Z{})$, as shown in 
Figure~\ref{fig:ModuliFundamentalDomain}. If we include the \CP transformation that descends from 
the transformation $M\mapsto -\overline{M}$, we obtain the full finite modular group 
$\mathrm{GL}(2,3)$, or $[48,29]$ according to the classification of GAP~\cite{GAP4}. It is a group 
with 48 elements (as the first entry in the bracket indicates).

\begin{figure}[h!]
\centering
\includegraphics[height=4.5in]{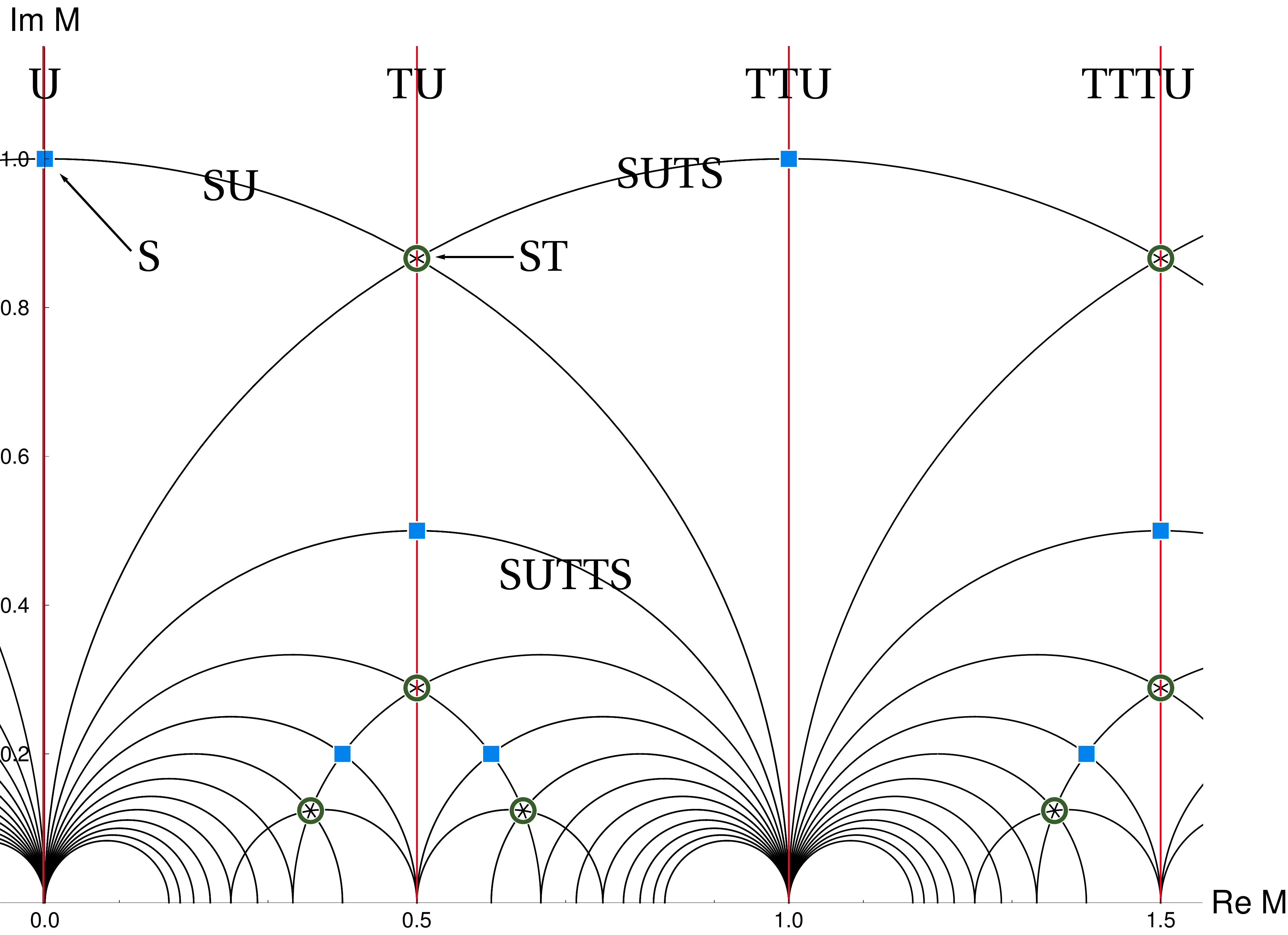}
\caption{Unbroken modular symmetries at various fixed points and fixed lines in moduli space, 
where $\mathrm{U}$ denotes the \CP-like transformation $M \mapsto -\overline{M}$.}
\label{fig:fixedPoints}
\end{figure}

In a full string construction that allows for predictions at low energies, the modulus $M$ must be 
stabilized. If the modulus $M$ is fixed at a generic point in moduli space, the modular symmetry 
breaks down completely. However, if the modulus adopts its value at a special fixed point (or fixed 
higher-dimensional sub-locus) in moduli space, some modular generators are unbroken and build a 
residual modular symmetry. If for example $M$ takes the value $\I$, only the modular subgroup 
generated by $\mathrm{S}$ is preserved, see Figure~\ref{fig:fixedPoints}.

\section{The eclectic flavor group}

We can now summarize the discrete flavor symmetries of string theory that can be determined via the 
outer automorphisms of the Narain space group. We have
\begin{itemize}
\item {\bf traditional flavor symmetries} that are universal in moduli space. In our example this 
corresponds to the flavor group $\Delta(54)$.

\item a subset of the {\bf finite modular group} that acts as a symmetry at specific fixed points 
(or fixed higher-dimensional sub-loci) in moduli space.  

\item at these ``points'' we have an {\bf enhanced flavor symmetry} that combines the traditional 
flavor symmetry with the unbroken elements of the modular symmetry group (in our example, the 
finite modular group $\mathrm{GL}(2,3)$).

\item in addition, matter fields have {\bf specific modular weights $k$}, which leads to further 
restrictions on the action of the theory (for example, on the Yukawa couplings). Moreover, they 
give rise to further enhancements of the symmetries in form of $R$-symmetries. 
\end{itemize}
The full flavor symmetry is thus non-universal in moduli space. At specific ``points'' we have an
enhancement of the traditional flavor symmetry (which  itself is universal in moduli space).

This brings us to the definition of the ``Eclectic Flavor Group''~\cite{Nilles:2020nnc}. Let us 
discuss this first in a simplified situation where we ignore the contributions of the automorphy 
factors that accompany the transformation of matter fields with a nontrivial weight. We then have 
to combine 
\begin{itemize}
\item the traditional flavor group ($\Delta(54)$ in our example) and
\item the finite modular flavor group that transforms the moduli as well (here $T^\prime$, or
$\mathrm{GL}(2,3)$ when \CP is included).
\end{itemize} 
The eclectic flavor group is now defined as the multiplicative closure of these groups. For the 
\Z{3} orbifold we obtain
\begin{itemize}
\item $\Omega(1) \cong [648,533]$ from $\Delta(54)$ and $T^\prime$,
\item $[1296,2891]$ from $\Delta(54)$ and $\mathrm{GL}(2,3)$, including \CP.
\end{itemize}
The eclectic flavor group is the largest possible flavor group for a given system, but it is not 
necessarily linearly realized. Part of it is spontaneously broken by the vacuum expectation values 
of the moduli. For generic values of the moduli only the traditional flavor symmetry remains 
unbroken. 

So far our simplified discussion. As we had already stressed earlier, the restrictions of modular 
symmetry are more severe than just represented by the finite modular flavor group (here $T^\prime$). 
In addition, we have to take into account the automorphy factors that come with the modular weights 
of the matter fields. These modular weights are determined in the specific string theory under 
consideration~\cite{Ibanez:1992hc,Olguin-Trejo:2017zav} and lead to an extension of the eclectic 
flavor group via additional $R$-symmetries. This has been discussed in detail in 
refs.~\cite{Nilles:2020tdp,Nilles:2020gvu}. For the \Z{3} orbifold, we can identify a symmetry 
$\Z{9}^R$ that extends $\Omega(1) \cong [648,533]$ to $\Omega(2) \cong [1944,3448]$ without 
\CP. If we further include \CP, we obtain a group with 3888 elements. 

\begin{figure}[h!]
\centering
\includegraphics[height=4.5in]{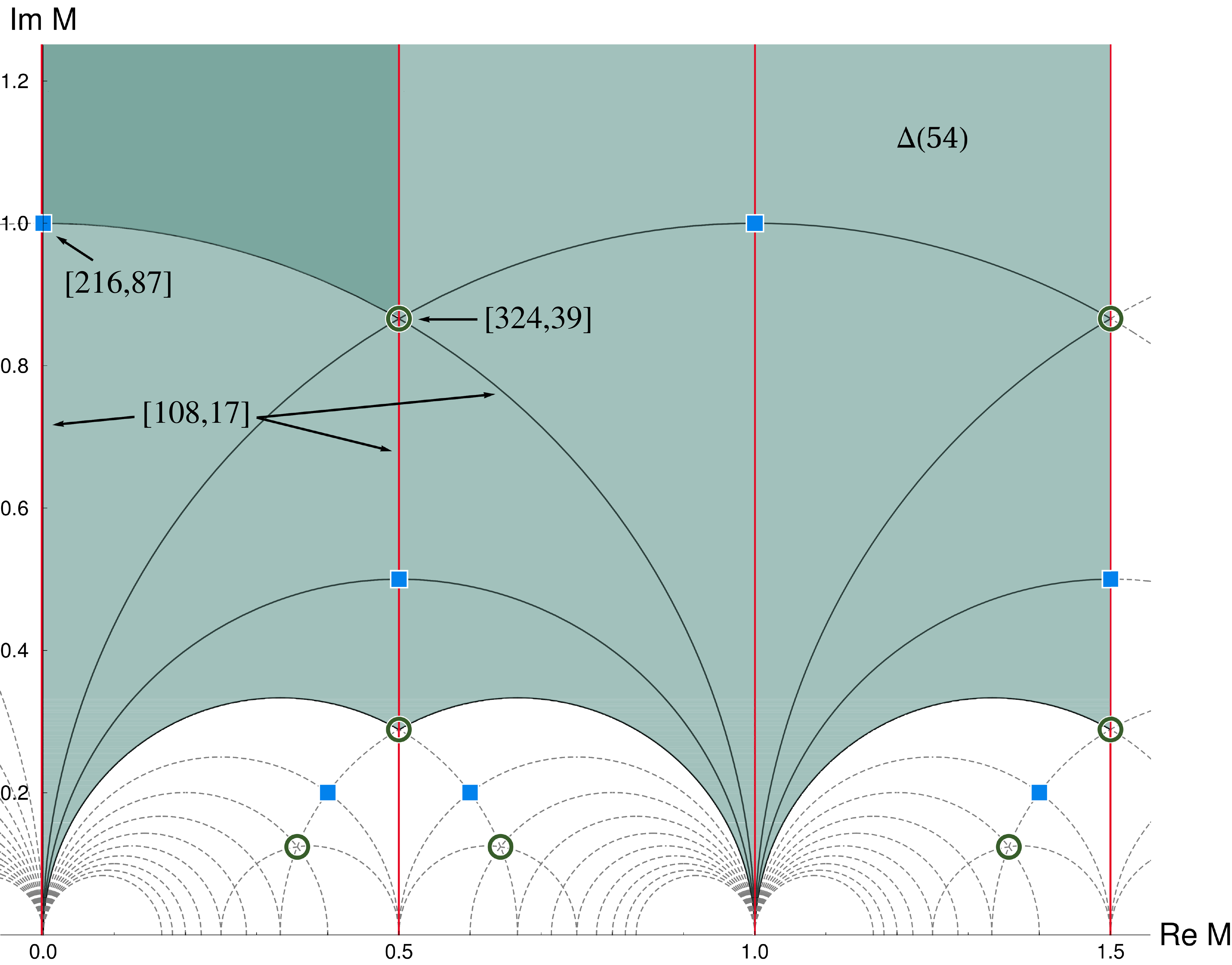}
\caption{Points and curves of the enhancement of flavor symmetries. On the vertical red lines and the black
semi-circles the traditional flavor group $\Delta(54)$ gets enhanced to [108,17]. Further enhancements
appear when two or three of these lines intersect.}
\label{fig:Z3Enhancements}
\end{figure}

\section{Modular flavor: top-down versus bottom-up}

\subsection{\bf Top-Down}

We have seen that the flavor symmetries in the top-down (TD) approach are very restrictive and lead 
to a scheme with high predictive power. Some of the reasons are:
\begin{itemize}
\item restrictions from the finite modular flavor group $\Gamma_N$ or its double cover $\Gamma_N'$,
\item specific representations for matter fields are selected and fixed by the underlying string 
symmetry,
\item these restrictions also hold for the values of the modular weights and the resulting 
``shaping symmetries'',
\item the role of flavon fields is partially played by the moduli, and
\item the modular flavor symmetries are always accompanied by a traditional flavor symmetry 
(with further restrictions on superpotential and K\"ahler potential of the theory~\cite{Nilles:2020kgo}).
\end{itemize}
We arrive at a large eclectic flavor group that reflects the symmetries of 
the underlying string theory and is not 
subject to arbitrary choices allowed in the bottom-up approach. 

Let us illustrate these facts in the framework of our \Z{3} orbifold example, as discussed in 
ref.~\cite{Nilles:2020kgo}. The twisted states at the three fixed points correspond to a triplet 
representation of $\Delta(54)$. Other states in the low energy sector correspond to singlets (both 
trivial and nontrivial). Winding states correspond to doublets of $\Delta(54)$ and are 
heavy~\cite{Nilles:2018wex}. The twisted states, however, do not transform as a triplet 
representation of the finite modular group $T^\prime$, but as a combination of a nontrivial 
doublet $\rep{2}^\prime$ and the trivial singlet $\rep{1}$. Other light states transform as 
singlets under $T^\prime$. The modular weights of the twisted fields are restricted to the 
values $k=\nicefrac{-2}{3}$ or $\nicefrac{-5}{3}$ (in the first twisted sector) and cannot be 
chosen freely (matter fields from the second twisted sector have $k=\nicefrac{-1}{3}$ and 
$\nicefrac{2}{3}$). All of these ingredients, the traditional flavor symmetry, the modular flavor 
symmetry and the specific automorphy factors lead to a very restrictive eclectic scheme.

\subsection{\bf Bottom-Up}

The application of modular symmetries for flavor physics has been pioneered in a remarkable paper 
of Feruglio~\cite{Feruglio:2017spp}. In his example he considers the lepton sector with finite 
modular flavor group $\Gamma_3 \cong A_4$. He assigns the left-handed leptons to the triplet 
representation of $A_4$ and the right-handed leptons to a combination 
$\rep{1} \oplus \rep{1}^\prime \oplus \rep{1}^{\prime\prime}$ of trivial and nontrivial singlets. 
Specific choices of the modular weights of matter fields lead to a successful fit for lepton masses 
and mixing angles. By now there have been many more model constructions based on other finite 
modular groups like $\Gamma_4$ or $\Gamma_5$ and their double covers~with and without \CP (see 
e.g.\ refs.~\cite{Kobayashi:2018vbk,Novichkov:2019sqv,Liu:2019khw,Gui-JunDing:2019wap,Novichkov:2020eep,Wang:2020lxk,Novichkov:2021evw}). 
Model building includes free choices of representations and modular weights of the matter fields. 
Further, admissible terms in the K\"ahler potential that may be phenomenologically relevant are 
typically disregarded~\cite{Chen:2019ewa}. In these constructions the presence of an additional 
traditional flavor symmetry is usually ignored, although it might further restrict the couplings of 
the theory and influence the results of the fit.

\subsection{\bf The Representation Dilemma}

Given the status of the field, at the moment a comparison of top-down (TD) and bottom-up (BU) 
constructions appears to be pretty difficult. First, we have to match the groups. While 
BU-approaches use a wide variety of finite modular flavor groups, the TD-approaches seem to be very 
restrictive. For example, the group $A_4$ is difficult to find. Observe that it is not a subgroup 
of its double cover $T^\prime$, which we find in the $\Z{3}$ orbifold. From that point of view, it 
might be more rewarding to intensify model building with modular flavor group $T^\prime$ (for 
recent discussions, see ref.~\cite{Liu:2019khw}).

The next step represents an even more serious challenge: the choice of the representations of the 
matter fields. Successful fits in the BU-approach typically assume a variety of representations of 
matter fields, like the triplets and various nontrivial singlets mentioned earlier. In the 
TD-approach, the choices seem to be much more limited. If we consider the model based on 
$T^\prime$, we see that the candidate matter fields do not transform as a triplet but as 
$\rep{1} \oplus \rep{2}^\prime$ and in addition we do not have all the necessary nontrivial 
singlets in the low energy spectrum. The same is true for the modular weights of the matter fields 
which are fixed in the TD-approach.

To match TD- and BU-approaches more work is needed. On the one hand we need more explicit 
TD-constructions to gain a better control of the relevant groups and representations. On the other 
hand one might try to consider BU-models with  more limited number of representations and modular 
weights. One would also try to understand the role of the traditional flavor symmetry that comes 
automatically in TD constructions and that has widely been neglected in modular BU models. In 
addition, the spontaneous breakdown of the traditional flavor symmetry has to be studied in 
detail. Much work remains to be done.

\section{Local flavor unification}

The eclectic flavor group appears as the multiplicative closure of the traditional flavor group 
and the (finite) modular symmetries. The former holds universally in moduli space while the 
latter are broken for generic values of the moduli. Nonetheless, for some specific values of the 
moduli (fixed sub-loci of the modular transformations), part of the modular symmetries are unbroken. 
This results in an enhancement of the traditional flavor group, leading to so-called ``Local 
Flavor Unification''. The full flavor symmetry is non-universal in moduli space and the spontaneous 
breakdown of modular flavor symmetry can be understood as a motion in moduli space. If we move away 
from these ``fixed points'', part of the flavor symmetry is broken. In the following we shall 
illustrate this mechanism within two specific examples: the $\Z{3}$ and the $\Z{2}$ orbifold. 

\subsection{$\Z{3}$ orbifold}

Here, we have the finite modular group $T^\prime$, being the double cover of $\Gamma_3 \cong A_4$. 
The fundamental domain of its moduli space is shown in Figure~\ref{fig:ModuliFundamentalDomain}. It 
reflects the restriction of the modular transformations  $\mathrm{S}$
and $\mathrm{T}$ (with the $\mod 3$ condition) on the modulus $M$. In addition, we have the 
\CP-like transformation
$$
\mathrm U: M ~\mapsto~ -\overline{M}\,,
$$
which reduces this fundamental domain by a factor of two (e.g.\ to the region 
$\mathrm{Re} M\geq 0$). This moduli space with its fixed points and fixed lines is shown in 
Figure~\ref{fig:fixedPoints}. It also shows the generators that act trivially on some of 
these fixed points and lines. The vertical red lines, for example, correspond to an unbroken \CP 
transformation. The fixed points of $\mathrm{SL}(2,\Z{})$ are at $M=\I$ (stabilized by the 
generator $\mathrm{S}$) and at $M=\exp(\pi\I/3)$ (stabilized by\ $\mathrm{S}\mathrm{T}$). 
At the fixed points and lines of Figure~\ref{fig:fixedPoints}, we obtain enhancements 
of the flavor group shown in Figure~\ref{fig:Z3Enhancements}. For generic values of the modulus $M$ 
we have the traditional flavor group $\Delta(54)$. On the lines we have an enhanced group with 108 
elements: [108,17]. At the points where two lines meet (blues squares) this group is further 
enhanced to [216,87]. The meeting points of three lines (green circles) leads to [324,39], the 
maximal enhancement in the present example. If one moves away from the fixed points (lines) the 
symmetries are (partially) spontaneously broken.

\subsection{$\Z{2}$ orbifold}

Up to now we have concentrated on the two-dimensional $\Z{3}$ orbifold. Let us now add the 
discussion of the $\Z{2}$ orbifold, as detailed in refs.~\cite{Baur:2020jwc,Baur:2021mtl}. It is of 
interest as it is the simplest example of an orbifold with both unconstrained (complex) moduli of 
the two-dimensional torus: the K\"ahler modulus (usually called $T$, not to be confused with the 
modular transformation $\mathrm{T}$) and the complex structure modulus ($U$). Then, modular 
transformations form a direct product group $\mathrm{SL}(2,\Z{})_T\times \mathrm{SL}(2,\Z{})_U$. 
This is in contrast to the \Z{3} orbifold, where the complex structure modulus $U=\exp(2\pi\I/3)$ 
needs to be fixed to allow for the \Z{3} twist, breaking $\mathrm{SL}(2,\Z{})_U$ to a remnant 
$R$-symmetry. In addition, the traditional flavor symmetry of the \Z{2} orbifold is
$$
(D_8\times D_8)/\Z2 ~\cong~ [32,49]
$$
as a result of the geometry and string selection rules. The two-dimensional {$\Z{2}$ orbifold can 
be understood as the combination of two twisted circles, each contributing $D_8$ to the traditional 
flavor group. The eclectic flavor group of the $\Z{2}$ orbifold is completed by the finite modular 
symmetry which contains the product $\Gamma_2^T\times \Gamma_2^U \cong S_3\times S_3$ originating 
from $\mathrm{SL}(2,\Z{})_T\times \mathrm{SL}(2,\Z{})_U$. In addition, the mirror symmetry, which 
exchanges the two unconstrained moduli $T$ and $U$, is manifest in the \Z2 orbifold. It acts as a 
\Z{2} symmetry on the moduli and as a \Z4 symmetry on the twisted string modes. This leads to 
the finite modular group
$$
[(S_3^T\times S_3^U)\rtimes \Z4^{\rm mirror}] \times \Z{2}^{\CP} ~\cong~ [288,880]
$$
including the \CP-like symmetry acting as $T\mapsto -\overline{T}$ and $U\mapsto -\overline{U}$ on 
the moduli. Inclusion of the automorphy factors (with fractional modular weights for matter fields) 
leads to an additional $\Z{4}^R$ $R$-symmetry. The eclectic flavor group is then the multiplicative 
closure of
$$
(D_8\times D_8)/\Z{2}\,,\quad (S_3^T\times S_3^U)\rtimes \Z4^{\rm mirror}\,,\qquad \Z4^R\quad \text{and}\quad \Z2^{\CP}\,,
$$
and has a total of 9216 elements.

Again, only part of the eclectic flavor symmetry is linearly realized. At generic values of the 
moduli we have the traditional flavor symmetry. It is enhanced at various fixed sub-loci of the 
modular flavor group [288,880] which includes $S_3^T$, $S_3^U$, $\Z{4}^{\rm mirror}$ and 
$\Z{2}^{\CP}$. The moduli space has now four real (two complex) dimensions as indicated in 
Figure~\ref{fig:Z2withGamma}. There we have defined the curves $\lambda_T$ and $\lambda_U$ as the 
boundaries of the fundamental domains of $\mathrm{SL}(2,\Z{})_T\times \mathrm{SL}(2,\Z{})_U$, 
respectively. Combined with the fixed domains of mirror symmetry ($T=U$) and the \CP-like symmetry 
we expect the enhanced unified flavor groups along these lines $\lambda_T$ and $\lambda_U$. The 
result of the full analysis is shown in Figure~\ref{fig:Z2FlavorLandscape}. The figure illustrates 
the intriguing interplay of mirror symmetry and finite modular symmetries and we see a variety of 
local unified flavor groups. The largest group is found at the point $T=U=\exp(\I\pi/3)$:
$$ 
([1152,157463]\times \Z4^R)/\Z2\;.
$$
A detailed discussion of Figure~\ref{fig:Z2FlavorLandscape} can be found in ref.~\cite{Baur:2021mtl}.

\begin{figure}[h!]
\centering
\includegraphics[height=3in]{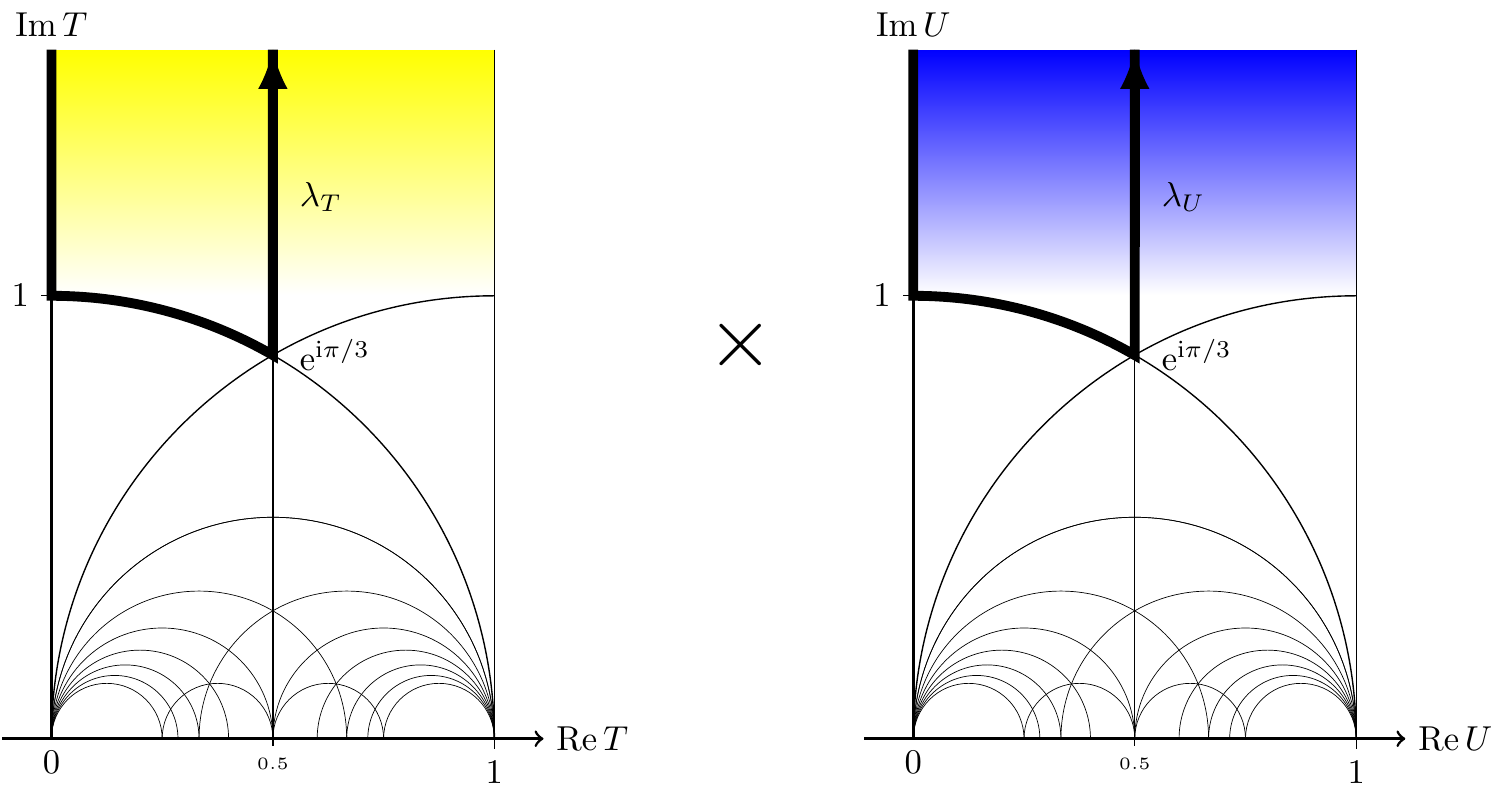}
\caption{The curves $\lambda_T$ and $\lambda_U$ are the boundaries of the corresponding fundamental 
domains. They start at $T=\I\infty$, pass by the fixed points and end at $T=0.5+\I\infty$. It 
is at these curves where we expect the enhancement of flavor groups.}
\label{fig:Z2withGamma}
\end{figure}

\begin{figure}[h!]
\centering
\includegraphics[height=6in]{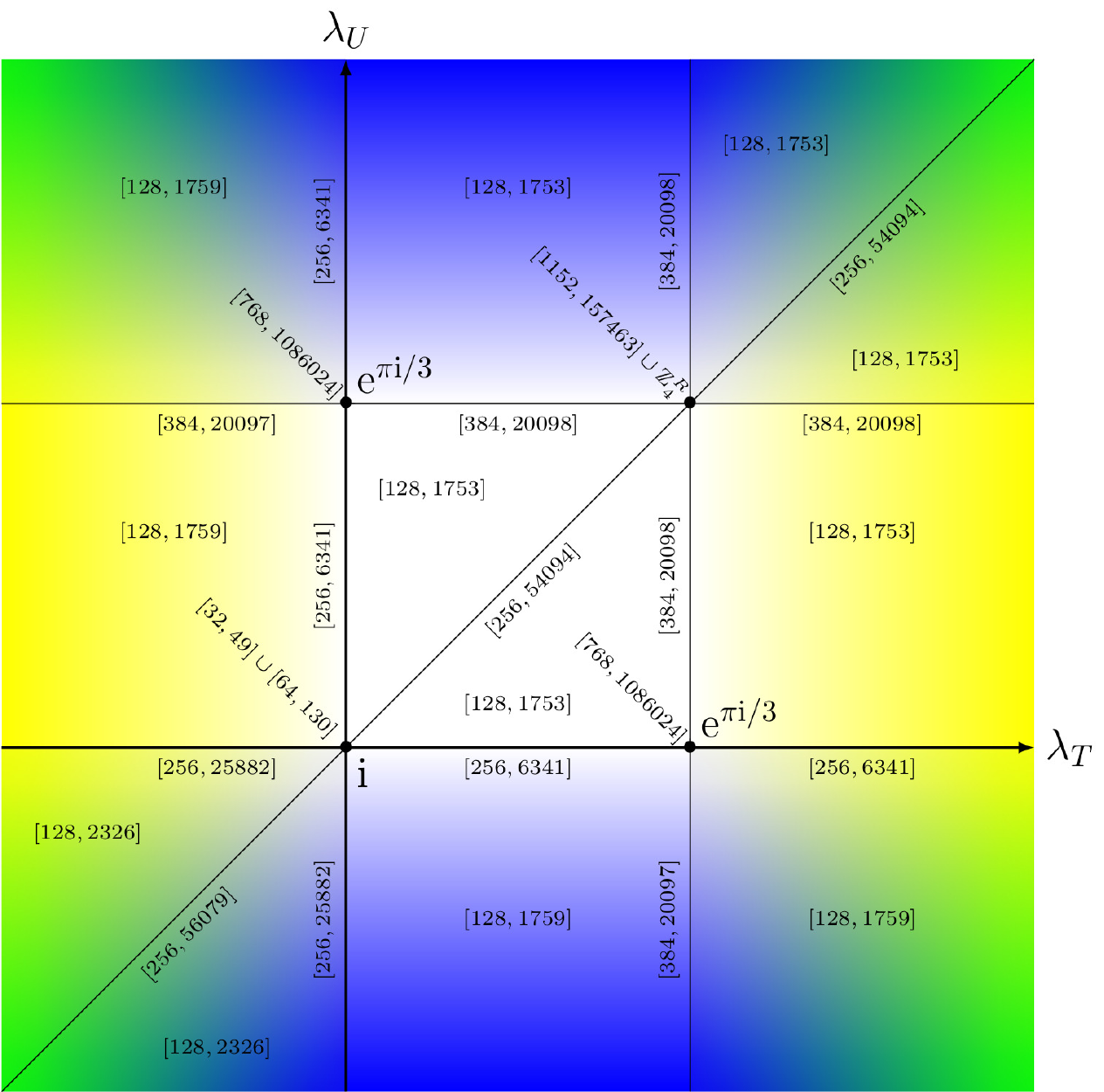}
\caption{Landscape of the unified flavor groups (including \CP) in the $\lambda_T$-$\lambda_U$ plane. 
We use the abbreviations $[32,49]\cup[64,130]\cong\big[[32,49]\times [64,130]\big]/\Z2$ and 
$[1152,157463]\cup\Z4^R\cong \left[[1152,157463]\times \Z4^R\right]/\Z2$, where the \Z2 corresponds
to the point group selection rule of the $\mathbbm T^2/\Z2$ orbifold sector, contained in both 
groups in the product. The axes $\lambda_T$ and $\lambda_U$ describe the boundaries of the 
fundamental domains of $T$ and $U$ (as shown in Figure~\ref{fig:Z2withGamma}). 
The diagonal line corresponds to  the hypersurface where $ U=T$ on the curves $\lambda_T$ and $\lambda_U$. 
The unified flavor groups above and below the diagonal are related by mirror symmetry $\overline M$.
}
\label{fig:Z2FlavorLandscape}
\end{figure}

\section{Wilson lines and $\mathrm{Sp}(4,\Z{})$}

The interplay between mirror symmetry and finite modular symmetries can be made manifest with the 
consideration of the Siegel modular group $\mathrm{Sp}(4,\Z{})$. From the string theory point of 
view, it appears as the simplest manifestation of gauge background fields (Wilson 
lines) on the torus~\cite{Baur:2020yjl}. $\mathrm{Sp}(4,\Z{})$ describes three moduli ($T$, $U$ 
and a Wilson line modulus $Z$) and it contains $\mathrm{SL}(2,\Z{})_T\times\mathrm{SL}(2,\Z{})$ 
and \Z2 mirror symmetry as subgroups. For a toroidal string compactification, the generators of 
$\mathrm{Sp}(4,\Z{})$ can be determined via the standard analysis of the outer automorphisms of 
the Narain space group. An application to the flavor structure of particle physics again requires 
orbifold twists to obtain chiral fermions. All possible twists are connected to the fixed sub-loci 
in the fundamental domain (i.e.\ the Siegel upper half plane) of $\mathrm{Sp}(4,\Z{})$~\cite{Nilles:2021pr}. 
They include the previously considered $\Z{K}$ orbifolds ($K=2,3,4,6$). In addition, a twist by 
\Z2 mirror symmetry is included and this twist leads to asymmetric orbifolds, where 
$\theta_\mathrm{L}\neq\theta_\mathrm{R}$. From the top-down approach, applications to flavor 
symmetries have not yet been discussed in detail.

Up to now, there has been one bottom-up attempt of flavor model building with 
$\mathrm{Sp}(4,\Z{})$, where the maximal finite modular symmetry was chosen as $S_6$~\cite{Ding:2020zxw}. 
In the case of vanishing third modulus (the Wilson line), $S_6$ includes and unifies the finite 
modular groups $(S_3\times S_3)\rtimes \Z2$ that appeared earlier in the discussion of the $\Z2$ 
orbifold (where mirror symmetry acts as $\Z2$ on the moduli). Hence, also $\mathrm{Sp}(4,\Z{})$ is 
expected to include, among others, a generalization of the \Z2 orbifold. The investigation of the 
Siegel modular group towards flavor symmetry has just started and we can look forward to many new 
and exciting aspects of flavor physics. One of them has been found already in the natural appearance 
of a \CP-like symmetry that extends $\mathrm{Sp}(4,\Z{})$ to $\mathrm{GSp}(4,\Z{})$.

\section{Where are we?}

The top-down approach to flavor symmetries predicts the presence of (discrete) traditional and 
modular flavor symmetries. It unifies these symmetries including \CP within the framework of the 
eclectic flavor group. While the traditional flavor symmetry is universal in moduli space, there 
appear non-universal enhancements of flavor symmetries and \CP at specific ``points'' in moduli 
space. The spontaneous breakdown of these symmetries can be understood as a motion in moduli space 
(away from these points or lines of enhancement). If we are close to these ``points'', the enhanced 
symmetries are only slightly broken. The breakdown of the traditional flavor symmetry, however, 
requires the presence of flavon fields.

This opens up a new arena for flavor model building to be compared to the existing bottom-up 
constructions. We need more explicit string theory constructions to exhaust the possibilities and 
to clarify the restrictions. These restrictions do not only concern the possible discrete groups 
but also the specific representations and modular weights of the matter fields in the low energy 
sector. At the moment, there is still a huge gap between existing top-down and bottom-up attempts.

As we have seen, string theory includes all the necessary ingredients for a discussion of flavor 
symmetries. These include:
\begin{itemize}

\item {\bf the traditional flavor group,}

\item {\bf the finite modular flavor group,}

\item {\bf additional $R$-symmetries (shaping symmetries) from the restrictions of the automorphy 
factors and the modular weights of matter fields, and}

\item {\bf a natural candidate for a \CP-symmetry,}

\end{itemize}
such that the eclectic flavor group serves as a unified description of the discrete flavor scheme.


\begin{thebibliography}{10}

\bibitem{Feruglio:2019ktm}
F.~Feruglio and A.~Romanino, \emph{{Lepton Flavour Symmetries}},  (2019),
  \texttt{arXiv:1912.06028} [hep-ph].

\bibitem{Ishimori:2010au}
H.~Ishimori, T.~Kobayashi, H.~Ohki, Y.~Shimizu, H.~Okada, and M.~Tanimoto,
  \emph{{Non-Abelian Discrete Symmetries in Particle Physics}}, Prog. Theor.
  Phys. Suppl. \textbf{183} (2010), 1--163, \texttt{arXiv:1003.3552} [hep-th].

\bibitem{Kobayashi:2006wq}
T.~Kobayashi, H.~P. Nilles, F.~Pl{\"o}ger, S.~Raby, and M.~Ratz, \emph{{Stringy
  origin of non-Abelian discrete flavor symmetries}}, Nucl. Phys. B
  \textbf{768} (2007), 135--156, \texttt{hep-ph/0611020}.

\bibitem{Dixon:1985jw}
L.~J. Dixon, J.~A. Harvey, C.~Vafa, and E.~Witten, \emph{{Strings on
  Orbifolds}}, Nucl. Phys. B \textbf{261} (1985), 678--686.

\bibitem{Dixon:1986jc}
L.~J. Dixon, J.~A. Harvey, C.~Vafa, and E.~Witten, \emph{{Strings on Orbifolds.
  2.}}, Nucl. Phys. B \textbf{274} (1986), 285--314.

\bibitem{Ibanez:1986tp}
L.~E. Ib{\'a}{\~n}ez, H.~P. Nilles, and F.~Quevedo, \emph{{Orbifolds and Wilson
  Lines}}, Phys. Lett. B \textbf{187} (1987), 25--32.

\bibitem{Buchmuller:2005jr}
W.~Buchm{\"u}ller, K.~Hamaguchi, O.~Lebedev, and M.~Ratz, \emph{Supersymmetric
  standard model from the heterotic string}, Phys. Rev. Lett. \textbf{96}
  (2006), 121602, \texttt{hep-ph/0511035}.

\bibitem{Lebedev:2006kn}
O.~Lebedev, H.~P. Nilles, S.~Raby, S.~Ramos-S{\'a}nchez, M.~Ratz, P.~K.~S.
  Vaudrevange, and A.~Wingerter, \emph{A mini-landscape of exact {MSSM} spectra
  in heterotic orbifolds}, Phys. Lett. \textbf{B645} (2007), 88,
  \texttt{hep-th/0611095}.

\bibitem{Nilles:2008gq}
H.~P. Nilles, S.~Ramos-S\'anchez, M.~Ratz, and P.~K.~S. Vaudrevange,
  \emph{{From strings to the MSSM}}, Eur. Phys. J. \textbf{C59} (2009),
  249--267, \texttt{arXiv:0806.3905} [hep-th].

\bibitem{Olguin-Trejo:2018wpw}
Y.~Olgu{\'i}n-Trejo, R.~P{\'e}rez-Mart{\'i}nez, and S.~Ramos-S{\'a}nchez,
  \emph{{Charting the flavor landscape of MSSM-like Abelian heterotic
  orbifolds}}, Phys. Rev. \textbf{D98} (2018), no.~10, 106020,
  \texttt{arXiv:1808.06622} [hep-th].

\bibitem{Baur:2019kwi}
A.~Baur, H.~P. Nilles, A.~Trautner, and P.~K.~S. Vaudrevange,
  \emph{{Unification of Flavor, CP, and Modular Symmetries}}, Phys. Lett. B
  \textbf{795} (2019), 7--14, \texttt{arXiv:1901.03251} [hep-th].

\bibitem{Carballo-Perez:2016ooy}
B.~Carballo-P{\'e}rez, E.~Peinado, and S.~Ramos-S{\'a}nchez,
  \emph{{$\Delta(54)$ flavor phenomenology and strings}}, JHEP \textbf{12}
  (2016), 131, \texttt{arXiv:1607.06812} [hep-ph].

\bibitem{Baur:2019iai}
A.~Baur, H.~P. Nilles, A.~Trautner, and P.~K.~S. Vaudrevange, \emph{{A String
  Theory of Flavor and $\mathcal{CP}$}}, Nucl. Phys. B \textbf{947} (2019),
  114737, \texttt{arXiv:1908.00805} [hep-th].

\bibitem{Narain:1985jj}
K.~S. Narain, \emph{{New Heterotic String Theories in Uncompactified Dimensions
  \ensuremath{<} 10}}, Phys. Lett. B \textbf{169} (1986), 41--46.

\bibitem{Narain:1986am}
K.~S. Narain, M.~H. Sarmadi, and E.~Witten, \emph{{A Note on Toroidal
  Compactification of Heterotic String Theory}}, Nucl. Phys. B \textbf{279}
  (1987), 369--379.

\bibitem{Narain:1986qm}
K.~S. Narain, M.~H. Sarmadi, and C.~Vafa, \emph{{Asymmetric Orbifolds}}, Nucl.
  Phys. B \textbf{288} (1987), 551.

\bibitem{GrootNibbelink:2017usl}
S.~Groot~Nibbelink and P.~K.~S. Vaudrevange, \emph{{T-duality orbifolds of
  heterotic Narain compactifications}}, JHEP \textbf{04} (2017), 030,
  \texttt{arXiv:1703.05323} [hep-th].

\bibitem{GAP4}
The GAP~Group, \emph{{GAP -- Groups, Algorithms, and Programming, Version
  4.11.1}}, 2021, \texttt{\url{https://www.gap-system.org}}.

\bibitem{Nilles:2020nnc}
H.~P. Nilles, S.~Ramos-S{\'a}nchez, and P.~K.~S. Vaudrevange, \emph{{Eclectic
  Flavor Groups}}, JHEP \textbf{02} (2020), 045, \texttt{arXiv:2001.01736}
  [hep-ph].

\bibitem{Ibanez:1992hc}
L.~E. Ib{\'a}{\~n}ez and D.~L{\"u}st, \emph{{Duality anomaly cancellation,
  minimal string unification and the effective low-energy Lagrangian of 4-D
  strings}}, Nucl. Phys. B \textbf{382} (1992), 305--361,
  \texttt{hep-th/9202046}.

\bibitem{Olguin-Trejo:2017zav}
Y.~Olgu{\'i}n-Trejo and S.~Ramos-S{\'a}nchez, \emph{{K{\"a}hler potential of
  heterotic orbifolds with multiple K{\"a}hler moduli}}, J. Phys. Conf. Ser.
  \textbf{912} (2017), no.~1, 012029, \texttt{arXiv:1707.09966} [hep-th].

\bibitem{Nilles:2020tdp}
H.~P. Nilles, S.~Ramos-S\'anchez, and P.~K.~S. Vaudrevange, \emph{{Eclectic
  flavor scheme from ten-dimensional string theory -- I. Basic results}}, Phys.
  Lett. B \textbf{808} (2020), 135615, \texttt{arXiv:2006.03059} [hep-th].

\bibitem{Nilles:2020gvu}
H.~P. Nilles, S.~Ramos-S\'anchez, and P.~K.~S. Vaudrevange, \emph{{Eclectic
  flavor scheme from ten-dimensional string theory -- II. Detailed technical
  analysis}}, Nucl. Phys. B \textbf{966} (2021), 115367,
  \texttt{arXiv:2010.13798} [hep-th].

\bibitem{Nilles:2020kgo}
H.~P. Nilles, S.~Ramos-{S\'anchez}, and P.~K.~S. Vaudrevange, \emph{{Lessons
  from eclectic flavor symmetries}}, Nucl. Phys. B \textbf{957} (2020), 115098,
  \texttt{arXiv:2004.05200} [hep-ph].

\bibitem{Nilles:2018wex}
H.~P. Nilles, M.~Ratz, A.~Trautner, and P.~K.~S. Vaudrevange,
  \emph{{$\mathcal{CP}$ Violation from String Theory}}, Phys. Lett.
  \textbf{B786} (2018), 283--287, \texttt{arXiv:1808.07060} [hep-th].

\bibitem{Feruglio:2017spp}
F.~Feruglio, \emph{{Are neutrino masses modular forms?}}, From My Vast
  Repertoire ...: Guido Altarelli's Legacy (A.~Levy, S.~Forte, and G.~Ridolfi,
  eds.), 2019, {\texttt{arXiv:1706.08749}} [hep-ph], pp.~227--266.

\bibitem{Kobayashi:2018vbk}
T.~Kobayashi, K.~Tanaka, and T.~H. Tatsuishi, \emph{{Neutrino mixing from
  finite modular groups}}, Phys. Rev. \textbf{D98} (2018), no.~1, 016004,
  \texttt{arXiv:1803.10391} [hep-ph].

\bibitem{Novichkov:2019sqv}
P.~P. Novichkov, J.~T. Penedo, S.~T. Petcov, and A.~V. Titov,
  \emph{{Generalised CP Symmetry in Modular-Invariant Models of Flavour}}, JHEP
  \textbf{07} (2019), 165, \texttt{arXiv:1905.11970} [hep-ph].

\bibitem{Liu:2019khw}
X.-G. Liu and G.-J. Ding, \emph{{Neutrino Masses and Mixing from Double
  Covering of Finite Modular Groups}}, JHEP \textbf{08} (2019), 134,
  \texttt{arXiv:1907.01488} [hep-ph].

\bibitem{Gui-JunDing:2019wap}
G.-J. Ding, S.~F. King, X.-G. Liu, and J.-N. Lu, \emph{{Modular $S_{4}$ and
  $A_{4}$ symmetries and their fixed points: new predictive examples of lepton
  mixing}}, JHEP \textbf{12} (2019), 030, \texttt{arXiv:1910.03460} [hep-ph].

\bibitem{Novichkov:2020eep}
P.~P. Novichkov, J.~T. Penedo, and S.~T. Petcov, \emph{{Double cover of modular
  $S_4$ for flavour model building}}, Nucl. Phys. B \textbf{963} (2021),
  115301, \texttt{arXiv:2006.03058} [hep-ph].

\bibitem{Wang:2020lxk}
X.~Wang, B.~Yu, and S.~Zhou, \emph{{Double covering of the modular $A_5$ group
  and lepton flavor mixing in the minimal seesaw model}}, Phys. Rev. D
  \textbf{103} (2021), no.~7, 076005, \texttt{arXiv:2010.10159} [hep-ph].

\bibitem{Novichkov:2021evw}
P.~P. Novichkov, J.~T. Penedo, and S.~T. Petcov, \emph{{Fermion Mass
  Hierarchies, Large Lepton Mixing and Residual Modular Symmetries}},  (2021),
  \texttt{arXiv:2102.07488} [hep-ph].

\bibitem{Chen:2019ewa}
M.-C. Chen, S.~Ramos-{S\'anchez}, and M.~Ratz, \emph{{A note on the predictions
  of models with modular flavor symmetries}}, Phys. Lett. B \textbf{801}
  (2020), 135153, \texttt{arXiv:1909.06910} [hep-ph].

\bibitem{Baur:2020jwc}
A.~Baur, M.~Kade, H.~P. Nilles, S.~Ramos-S{\'a}nchez, and P.~K.~S. Vaudrevange,
  \emph{{The eclectic flavor symmetry of the $\mathbb{Z}_2$ orbifold}}, JHEP
  \textbf{02} (2021), 018, \texttt{arXiv:2008.07534} [hep-th].

\bibitem{Baur:2021mtl}
A.~Baur, M.~Kade, H.~P. Nilles, S.~Ramos-S{\'a}nchez, and P.~K.~S. Vaudrevange,
  \emph{{Completing the eclectic flavor scheme of the ${\mathbb Z_2}$
  orbifold}},  (2021), \texttt{arXiv:2104.03981} [hep-th].

\bibitem{Baur:2020yjl}
A.~Baur, M.~Kade, H.~P. Nilles, S.~Ramos-S\'anchez, and P.~K.~S. Vaudrevange,
  \emph{{Siegel modular flavor group and CP from string theory}}, Phys. Lett. B
  \textbf{816} (2021), 136176, \texttt{arXiv:2012.09586} [hep-th].

\bibitem{Nilles:2021pr}
H.~P. Nilles, S.~Ramos-S{\'a}nchez, A.~Trautner, and P.~K.~S. Vaudrevange,
  (2021), {\emph{in preparation}}.

\bibitem{Ding:2020zxw}
G.-J. Ding, F.~Feruglio, and X.-G. Liu, \emph{{Automorphic Forms and Fermion
  Masses}}, JHEP \textbf{01} (2021), 037, \texttt{arXiv:2010.07952} [hep-th].

\end{thebibliography}
\providecommand{\bysame}{\leavevmode\hbox to3em{\hrulefill}\thinspace}

\end{document}